\documentclass[sn-mathphys-num]{sn-jnl}

\usepackage{graphicx}%
\usepackage{multirow}%
\usepackage{amsmath,amssymb,amsfonts}%
\usepackage{amsthm}%
\usepackage{mathrsfs}%
\usepackage[title]{appendix}%
\usepackage{xcolor}%
\usepackage{textcomp}%
\usepackage{manyfoot}%
\usepackage{booktabs}%
\usepackage{algorithm}%
\usepackage{algorithmicx}%
\usepackage{algpseudocode}%
\usepackage{listings}%

\usepackage{orcidlink}

\newcommand{\pdrv}[2]{\frac{\partial #1}{\partial #2}}

\newcommand{\eps}{\varepsilon}

\newcommand{\ovl}[1]{{\overline{#1}}}

\newcommand{\op}{\operatorname}

\theoremstyle{plain}%
\newtheorem*{claim*}{Claim}

\theoremstyle{remark}%
\newtheorem*{remark*}{Remark}%

\theoremstyle{definition}%
\newtheorem{definition}{Definition}%

\begin{document}

\title[Convergence to Bohmian mechanics in a de Broglie-like pilot-wave system]{Convergence to Bohmian mechanics in a de Broglie-like pilot-wave system}


\author*{\fnm{David} \sur{Darrow} \orcidlink{0000-0003-2787-961X}}\email{ddarrow@mit.edu}



\affil*{\orgname{Massachusetts Institute of Technology}, \orgdiv{Department of Mathematics}, \city{Cambridge}, \state{MA}, \country{USA}}




\abstract{
	Bohmian mechanics supplements the quantum wavefunction with deterministic particle trajectories, offering an alternate, dynamical language for quantum theory. However, the Bohmian particle does not affect its guiding wave, so the {wave field} must instead be prescribed by the system geometry. While this property {is widely assumed} necessary to ensure agreement with quantum mechanics, much work has recently been dedicated to understanding classical pilot-wave systems, which feature a two-way coupling between particle and wave. These systems---including the ``walking droplet'' system of Couder and Fort~\cite{PhysRevLett.97.154101} and its various abstractions~\cite{Durey_GPWF,hqfti_i,hqfti_ii,darrow_AM_model}---allow us to investigate the limits of classical systems and offer a touchstone between quantum and classical dynamics. 
	In this work, we present a general result that bridges Bohmian mechanics with this classical pilot-wave theory. Namely, Darrow and Bush~\cite{darrow_AM_model} recently introduced a Lagrangian pilot-wave framework to study quantum-like behaviours in classical systems; with a particular choice of particle-wave coupling, they recover {key dynamics} hypothesised in de Broglie's early \emph{double-solution} theory~\cite{deBroglie1956,deBroglie1970}.
	We {here} show that, with a different choice of coupling, {their de Broglie-like system} reduces exactly to single-particle Bohmian mechanics in the non-relativistic limit. 
	Finally, we present an application of the present work in developing an analogue for position measurement in a de Broglie-like setting.
}




\maketitle

\section{Introduction}
In his 1924 thesis~\cite{1925AnPh...10...22D}, Louis de Broglie predicted that massive particles would behave like waves of the {eponymous} \emph{de Broglie} wavenumber:
\[k = p/\hbar,\]
where $p = \gamma m u$ is the relativistic momentum of a particle of mass $m$ and velocity $u$. His vision deviated somewhat from the prevailing theories of Schr\"odinger, Bohr, and Heisenberg, however; in the subsequent years, he worked on his \emph{double-solution} theory, which posited that a quantum particle {consists} of \emph{both} particle and wave, evolving in unison~\cite{deBroglie1956,deBroglie1970}. The wave field would contain a singularity at the particle's position, though its form was {unspecified} by de Broglie, and the interplay between the particle and wave would give rise to {emergent} statistical behaviour in line with quantum theory.
De Broglie famously presented a version of his theory at the 1927 Solvay conference, to mixed reception, and {(temporarily)} abandoned the pilot-wave approach shortly thereafter~\cite{bacciagaluppi2009quantum}.

{In 1952,} David Bohm arrived at a different pilot-wave model of quantum mechanics~\cite{Bohm1952a,Bohm1952b}, providing an important counterexample to the ``impossibility proofs'' {of von Neumann and others, which had suggested the impossibility of any hidden variable theory}~\cite{30573279-e8ec-3f1e-b0ba-eaf73275f821}. Alongside the Schr\"odinger wavefunction $\psi$, Bohm posits a deterministic, physical trajectory for each particle in the system~\cite{Bohm1952a,Bohm1952b,Holland1993}. Restricting to the non-relativistic, one-particle case for simplicity, the wavefunction remains that of Schr\"odinger,
\begin{equation}\label{eq:bohm_psi}
	i\hbar\partial_t\psi = -\frac{\hbar^2}{2m}\nabla^2\psi + V(x)\psi,
\end{equation}
and the particle position $q_p$ is guided along as
\begin{equation}\label{eq:bohm_qp}
	\dot{q}_p = \frac{\hbar}{m}\op{Im}\left(\frac{\nabla\psi}{\psi}\right).
\end{equation}
Notably, if Born's rule (i.e., $P(q_p = q)\propto |\psi(q)|^2$) holds at one time, the trajectory equation (\ref{eq:bohm_qp}) ensures that it holds at arbitrary times---as such, {Bohmian} predictions align exactly with those of standard quantum theory~\cite{Bohm1952a}. 

{Bohm's work inspired de Broglie to return to pilot-wave dynamics~\cite{422ddf4d-944e-34dd-bb65-095bdb4bb08e}, and it also} caught the attention of John Bell, who identified the mathematical gaps in von Neumann's proof~\cite{RevModPhys.38.447}. {Bell also proved his own no-go theorem~\cite{PhysicsPhysiqueFizika.1.195}, demonstrating that any hidden variable approach to quantum mechanics must be inherently non-local; his work was corroborated by a famous experiment of Alain Aspect in 1982~\cite{PhysRevLett.49.1804}, lending credence to Bohm's non-local theory of particle trajectories.}

A key property of the Bohmian approach, however, is that the particle's Bohmian position has \emph{no influence} on the wavefunction. The wavefunction evolution is prescribed by the system geometry, and the particle navigates its guiding wave without altering it. While this property has thus far been assumed necessary for agreement with quantum mechanics (with a few exceptions, e.g., \cite{sutherland2019incorporating,sutherland2019incorporating2}), it marks a significant departure both from de Broglie's earlier pilot-wave approach and from classical field theories, and it raises real consequences for how one can interpret the theory. As an example, one can see the one-way coupling as equivalent to the problem of ``empty waves'' in Bohmian mechanics; if the particle cannot affect the wave field, it is difficult to distinguish a field that \emph{is} carrying a particle from one that is not.

In a different direction, there has been a great deal of recent interest in understanding \emph{classical} pilot-wave systems, largely inspired by the ``walking droplet'' system of Couder and Fort~\cite{PhysRevLett.97.154101}. In their system, a millimetric droplet moves about the surface of a vibrating fluid by interacting with a wave of its own creation. Walking droplets have been shown to engage in a variety of quantum-like behaviours: among many other examples~\cite{hqa}, they have been found to undergo single-particle slit diffraction and interference~\cite{PhysRevLett.97.154101,Pucci2018,ellegaard_interaction_2020}, attain quantised orbits and stable spin states~\cite{fort_2010,harris_bush_2014,perrard_2014,spinstates}, give rise to quantum-like statistics in an analogue of the quantum corral~\cite{PhysRevE.88.011001}, and follow ``surreal trajectories'' similar to those predicted by Bohmian mechanics~\cite{PhysRevA.106.L010203}.
While the majority of these analogues are only qualitative, recent authors have attempted to construct classical field theories to achieve quantitative agreement with certain quantum results \cite{hqfti_i,hqfti_ii}, furthering the theoretical program initiated by Louis de Broglie~\cite{deBroglie1956,deBroglie1970}. 

{In fact, like de Broglie's double-solution program before them, these classical systems} are defined by their two-way coupling; the highly nonlinear interplay between particle and field gives rise to interesting dynamical behaviours, seemingly intractably at odds with the simple wave-guiding-particle approach of Bohmian mechanics.

In the present work, we {work to connect} these two classes of pilot-wave theory. {We make use} of the Lagrangian pilot-wave framework put forward by Darrow and Bush~\cite{darrow_AM_model}, which couples a Klein--Gordon field to a relativistic point particle. With a particular choice of particle-wave coupling, they recover {many of the key behaviours hypothesised} in de Broglie's double-solution theory, as we describe in Section~\ref{sec:dB}.

We {investigate} {their framework with a different choice of coupling, recovering a distinct set of quantum-like behaviours. Namely,} we show that, in the non-relativistic limit, the system converges exactly to single-particle Bohmian mechanics, where the guiding wave (less a particle-centred wavepacket) plays the role of the Schr\"odinger wavefunction. {This result holds in a highly general setting of the pilot-wave framework, with arbitrary domain, initial conditions, and potential. Moreover, it} extends the relation between Klein--Gordon and Schr\"odinger waves~\cite{SCHOENE197936} to the case of a particle-wave system, and it {provides} a means by which Bohmian mechanics can emerge from a pilot-wave theory closer to de Broglie's original double-solution program.

A {related} program was {considered} by Sutherland~\cite{sutherland2019incorporating,sutherland2019incorporating2}, {who} showed that a certain particle-wave coupling could be ``gauged away'' to reveal a match {with} Bohmian mechanics. However, the current work undertakes a different mechanism to decouple the wave field from particle feedback, and which cannot be added or removed with an appropriate gauge transformation. This difference ensures that, in limits where it \emph{does not} converge to Bohmian mechanics, our system is able to undergo the same nonlinear evolution characteristic of the classical pilot-wave theories discussed above.

{As a demonstration of the latter, we introduce} an analogue of wavefunction collapse in our de Broglie-like pilot-wave system. Under an appropriately-defined analogue of a (single-particle) position measurement, we show that the wavefunction collapses exactly to its projection in the measured region containing the particle. 

The present work helps to bridge the gap between two different perspectives on pilot-wave dynamics. Bohmian mechanics exactly recovers the predictions of quantum mechanics, but at the cost of a one-way coupling {from wave to particle}; classical pilot-wave systems maintain a two-way coupling, and they can recover convincing analogues of quantum dynamics in certain cases, but it is far from clear whether they could replicate even single-particle quantum predictions completely. Our work {answers this latter question in the affirmative, and it} demonstrates how the pilot-wave framework of Darrow and Bush neatly interpolates between Bohmian mechanics and classical---or at least de Broglie-like---dynamics, suggesting a stronger link between these systems than previously thought. 

\section{De Broglie-like pilot-waves}\label{sec:dB}
We briefly review the Lagrangian pilot-wave framework put forward by Darrow and Bush~\cite{darrow_AM_model}, before moving on to the specific limit under consideration. Using a joint particle-wave action, they couple a complex Klein--Gordon field $\phi$ to a relativistic particle of mass $m$ at the point $\vec{q}_p\in\mathbb{R}^3$. Extremising against both objects, they find that the particle trajectory and field satisfy the coupled equations
\begin{equation}\label{eq:oureqs}
\begin{gathered}
	(\partial_\mu\partial^\mu + k_c^2)\phi = k_c^{-1}\gamma^{-1}\delta^3(\vec{q}-\vec{q}_p)\pdrv{\sigma}{\phi^*},\\
	d_t\left(m\sigma\gamma\vec{u}\right) = \gamma^{-1}mc^2\nabla\sigma(q_p),
\end{gathered}
\end{equation}
{where $\vec{u}$ is the velocity of the particle, $\gamma = (1 - (\frac{\vec{u}}{c})^2)^{-1/2}$ its Lorentz factor,} $k_c = mc/\hbar$ its Compton wavenumber, and $\sigma = \sigma(\phi)$ a real-valued coupling function\footnote{In fact, $\sigma$ should be a function of the continuous part $\ovl{\phi}$, which we define shortly. This distinction does not play into the current discussion.} between particle and wave. 

Darrow and Bush focus on a particular limit of interest, where
\[\sigma = 1 + b^2/4\pi + b\op{Re}\phi\]
for a real coupling parameter $b$. This limit formalises many of the properties de Broglie hypothesised for his \emph{double-solution} program~\cite{deBroglie1956,deBroglie1970}, including a synchronised Compton-scale oscillation in both particle and field, the de Broglie relation $\vec{p}=\hbar\vec{k}$ in the vicinity of the particle, and a Compton-scale wavepacket that traces out the particle's trajectory.


In the present work, we consider a new limit of (\ref{eq:oureqs}), where the particle is coupled to the wave's phase rather than its amplitude. Recall from Darrow and Bush~\cite{darrow_AM_model} the definition of the \emph{continuous component} of $\phi$:
\[\ovl{\phi}(q) = \lim_{r\to 0}\frac{1}{4\pi}\int_{S^2}d\xi\; \partial_r(r\phi(q + r\xi)).\]
Where $\phi$ is continuous, we find $\phi = \ovl{\phi}$; however, if 
\begin{equation}\label{eq:singpart}
\phi = \phi_1 + a/\|q-q_p\|
\end{equation}
for a continuous function $\phi_1$ near a point $q_p$, we instead find $\ovl{\phi}(q_p) = \phi_1(q_p)$. With this in mind, define {the continuous wave phase to be}
\[\theta := \op{Im}\log\ovl{\phi} = \frac{1}{2i}\left(\log\ovl{\phi} - \log\ovl{\phi}^*\right)\]
at the point $q_p$. This phase is well-defined (up to a period of $2\pi$) and finite.

Now, we define {our particle-wave coupling} by the following choice of sawtooth wave for $\sigma$: 
\begin{equation}\label{eq:funkyfunction}
\sigma(\phi) = a + b\theta, \qquad 0<\theta\leq 2\pi,
\end{equation}
where $\sigma/b < 0$ for all $\theta$; we can ensure this condition by choosing $a$ and $b$ to have opposite signs with $|a|>2\pi |b|$.

We make this choice of $\sigma$ to ensure that dynamics are independent of global phase shifts, but much of our analysis can be extended to a generic function $\sigma=\sigma(\theta)$. We will briefly comment on alternate choices in the following section.




Finally, applying (\ref{eq:oureqs}) to the coupling (\ref{eq:funkyfunction}), we find
\begin{equation}\label{eq:oureqsnow}
\begin{gathered}
	\left(\partial^\mu\partial_\mu + k_c^2\right)\phi = -\delta^3(q-q_p)\frac{b\gamma^{-1}}{2i k_c\ovl{\phi}^*},\\
	d_t(\gamma m\sigma(\theta)u) = -\gamma^{-1}mc^2\nabla \sigma(\theta).
\end{gathered}
\end{equation}
The means by which these equations reduce to (\ref{eq:bohm_psi}) and (\ref{eq:bohm_qp}) will be detailed in Section \ref{sec:dynamics}. 


There are several immediate differences we can identify from the limit discussed by Darrow and Bush~\cite{darrow_AM_model}. Two features are familiar from quantum mechanics: $\phi$ is now necessarily complex-valued, and the particle dynamics are now independent of field amplitude. For a more subtle distinction, recall that a Klein--Gordon field decomposes in the non-relativistic limit as $\phi\sim \psi e^{-i \omega_ct}$, taking on a global phase oscillation at the Compton frequency. In their limit, this oscillation imparts a particle vibration at the same frequency, as $\nabla\phi$ itself oscillates---this \emph{Zitterbewegung} plays a critical role in recovering aspects of de Broglie's double-solution theory. {With the current choice of coupling,} however, the force $\nabla\sigma \sim \nabla\theta$ no longer changes under a global change of phase, and we expect (and will demonstrate shortly) that the Zitterbewegung should disappear in the non-relativistic limit of {the present} system. 


\section{Reduction to Bohmian Mechanics}\label{sec:dynamics}
In parallel with the non-relativistic limit of the classical Klein--Gordon theory~\cite{SCHOENE197936}, we introduce an oscillating ansatz of the form
\begin{equation}\label{eq:ansatz}
\phi = \psi e^{-i\omega_c t} - \frac{be^{-i\omega_c t}}{8\pi i k_c \psi^*(q_p)\|q-q_p\|} =: \psi e^{-i\omega_c t} + \phi_\text{wav},
\end{equation}
with $\omega_c = mc^2/\hbar$ the Compton frequency of our particle, $k_c = mc/\hbar$ the corresponding wavenumber, and $b$ as in (\ref{eq:funkyfunction}). Notably, the continuous component of our wave is simply $\ovl{\phi} = \psi e^{-i\omega_c t}$. We show here that, {in the non-relativistic limit}, the evolution of $\psi$ approaches that of a Schr\"odinger wavefunction, and the evolution of $q_p$ approaches that of a Bohmian particle guided by $\psi$.

Specifically, we demonstrate the following result:

\bigskip
\begin{claim*}
Suppose $\phi$ and $q_p$ evolve under (\ref{eq:oureqsnow}) in a connected, open set $\Omega\subset\mathbb{R}^3$, with zero Dirichlet conditions on $\partial\Omega$ (if the boundary is non-empty).
Fix initial conditions $\psi_0$, $\dot{\psi}_0$, $q_{p,0}$, and $\dot{q}_{p,0}$, such that $\psi_0(q_{p,0})\neq 0$, and let $\phi$ be given by (\ref{eq:ansatz}). Let $(\psi^B,q_p^B)$ be the free Bohmian process in $\Omega$ with the same initial conditions.

In the non-relativistic limit $u\ll c$, $|\dot{\psi}|\ll \omega_c|\psi|$, the quantities $\psi$ and $q_p$ converge to $\psi^B$ and $q_p^B$, respectively.
\end{claim*}

\bigskip
\begin{remark*}
We present our argument only in the free (or wall-bounded) case for mathematical clarity. However, at the end of this section, we describe how {the same argument extends to a} non-zero potential $V(q)$.
\end{remark*}

\bigskip
To show this result, we instead demonstrate that such a convergence occurs for a \emph{short time}, with $t\in(0,\tau]$ for some
\[\tau=\tau(|\psi_0(q_{p,0})|,|\nabla\psi_0(q_{p,0})|,|\dot{q}_{p,0}|).\]
In particular, if we define
\[M = |\psi_0(q_{p,0})|^{-1}+|\nabla\psi_0(q_{p,0})|+|\dot{q}_{p,0}|,\]
we argue that $\tau$ is a strictly decreasing function of $M$. In Bohmian mechanics, however, all of these terms are locally uniformly bounded, since the particle never intersects the $\psi=0$ locus; that is, for any $T>0$, we can fix
\[M_T = \sup_{(0,T]}\left(|\psi^B(q^B_{p})|^{-1}+|\nabla\psi^B(q^B_{p})|+|\dot{q}^B_{p}|\right)\]
and choose a corresponding $\tau_T = \tau_T(M_T)$ that ensures convergence for any subprocess $t\in(t_0,t_0+\tau]\subset(0,T]$. Piecing these processes together, we see that the process itself converges over the full interval---and since $T$ was arbitrary, that it converges for all time.

\paragraph*{Trajectory equation.}
We focus first on the trajectory equation, as it allows us to rule out {fast particle oscillations}. Specifically, we show that, if the particle does not currently satisfy the guidance equation (\ref{eq:bohm_qp}), the ``error'' in its velocity is quickly dampened until it does. This property guarantees convergence to the Bohmian trajectory \emph{even} if the particle does not begin with the correct, Bohmian velocity. Moreover, this is a significant difference from the limit studied by Darrow and Bush~\cite{darrow_AM_model}, where the particle undergoes fast \emph{Zitterbewegung} oscillations about its mean trajectory.

As a note, because the singular component of $\phi$ is symmetric about $q_p$, we self-consistently define
\[\nabla\theta := \nabla\op{Im}\log(\psi e^{-i\omega_c t})\]
at the point $q_p$. At the level of derivatives, this is simply a mean gradient over a small ball centred at $q_p$.

The time derivative of $\sigma$ is
\begin{align*}
\dot{\sigma}(\theta) &= d_t(\op{Im}\log\psi - \omega_ct)\sigma'(\theta) \\
&= -\omega_c\sigma'(\theta)(1+O(|\dot{\psi}|/|\psi|\omega_c)),
\end{align*}
which transforms our guiding equation---to leading order in $u/c$---into
\[-\omega_c\sigma'(\theta)u + \sigma(\theta)\dot{u} = -c^2\sigma'(\theta)\nabla\theta + O(u^2/c^2).\]
Now, note that
\[\nabla\theta = \nabla \op{Im}\log\psi  = \op{Im}\left(\frac{\nabla\psi}{\psi}\right),\]
which allows us to recover
\[u - \frac{\sigma}{\omega_c\sigma'} \dot{u} = \frac{\hbar}{m}\op{Im}\left(\frac{\nabla\psi}{\psi}\right) + O(u^2/c^2),\]
or, plugging in our expression (\ref{eq:funkyfunction}),
\begin{equation}\label{eq:particle}
u - \frac{\sigma}{b\omega_c} \dot{u} = \frac{\hbar}{m}\op{Im}\left(\frac{\nabla\psi}{\psi}\right) + O(u^2/c^2).
\end{equation}
There are two possibilities with these dynamics.

Na\"ively, it appears that the $\dot{u}$ term vanishes in the non-relativistic limit, and this is indeed a solution of (\ref{eq:particle}). In this case, the dynamics reduce to
\begin{equation}\label{eq:particle_simple}
u = \frac{\hbar}{m}\op{Im}\left(\frac{\nabla\psi}{\psi}\right),
\end{equation}
identical to the equation (\ref{eq:bohm_qp}) guiding Bohmian particles.

As a second possibility, we can take a cue from the model of Darrow and Bush~\cite{darrow_AM_model} and note that these dynamics might support a transient motion \emph{about} the Bohmian trajectory (\ref{eq:particle_simple}). In particular, the full trajectory can generically take the form
\[u_\text{tot} = u_\text{Bohm} + u_\text{fluc},\]
where $u_\text{fluc}$ satisfies the unforced equation
\begin{equation}\label{eq:transient}
u_\text{fluc} - \frac{\sigma(-\omega_c t)}{b\omega_c}\dot{u}_\text{fluc} = 0.
\end{equation}
With the particle-wave coupling we have fixed, however, we have that $\sigma/b < 0$; then the equation (\ref{eq:transient}) dampens $u_\text{fluc}$ to zero within a Compton timescale $\sim 1/\omega_c$. 

\paragraph*{Field equation.} Following a similar argument as carried out by Darrow and Bush~\cite{darrow_AM_model} in their limit of interest, the ansatz (\ref{eq:ansatz}) ensures that $\psi$ is continuous at the point $q_p$, and that $\ovl{\phi}(q_p) = \psi(q_p)e^{-i\omega_c t}$. As such, we can write the nonlinear forcing in the wave equation as
\begin{equation}\label{eq:forcedwave}
\left(\partial^\mu\partial_\mu + k_c^2\right)\phi = -\delta^3(q-q_p)\frac{\gamma^{-1}be^{-i\omega_c t}}{2i k_c\psi^*}.
\end{equation}

To understand how the particle-to-wave and wave-to-particle effects might decouple in the non-relativistic limit, first consider a simple case where $q_p$ and $\psi$ are unchanging in time. Then the singular component of $\phi$ satisfies
\begin{equation}\label{eq:phiforcing}
(\partial^\mu\partial_\mu +k_c^2)\phi_\text{wav} = -\nabla^2 \phi_\text{wav} = -\delta^3(q-q_p)\frac{be^{-i\omega_ct}}{2i k_c\psi^*},
\end{equation}
which would exactly account for our nonlinear forcing. {As we will see, this simple case reflects the system in general; in the non-relativistic limit,} the Compton-frequency oscillation dominates the evolution of this source term, and the fixed-time solution $\phi_\text{wav}$ becomes a true solution to the forced Klein--Gordon equation. The remainder, $\psi e^{-i\omega_c t}$, {then evolves} according to the unforced equation.

More rigorously, the above argument identifies
\begin{equation}\label{eq:phi_fixed}
\phi_{\text{wav}, t_0} := -\frac{be^{-i\omega_c t}}{8\pi i k_c \psi^*(q_p,t_0)\|q-q_p\|}
\end{equation}
as the unique wavepacket solution to the fixed-$\psi$ problem
\[(\partial^\mu\partial_\mu + k_c^2)\phi_{\text{wav}, t_0} = -\delta^3(q-q_p)\frac{be^{-i\omega_c t}}{2i k_c\psi^*(q_p,t_0)}.\]
In moving to the generic case, where $\psi$ varies in time, we proceed to show that the inhomogeneous {portion of the} solution remains close to (\ref{eq:phi_fixed}).

To this end, recall the form of the retarded Klein--Gordon Green's function~\cite{wolfram_greens}:
\begin{align*}
G(q,t) &= \frac{\theta(t)}{4\pi\|q\|}\delta(t-\|q\|/c)\\
&\qquad-\frac{k_c^2}{2\pi}\theta(ct-\|q\|)\frac{J_1(\omega_c\sqrt{t^2-\|q\|^2/c^2})}{\omega_c\sqrt{t^2-\|q\|^2/c^2}}.
\end{align*}
Here, $\theta$ is the Heaviside step function, which ensures that data travels forward in time, and $J_1$ is a Bessel function of the first kind. The first term in this expression communicates a simple Coulomb potential; the second term corrects this potential with an internal vibration at the Compton frequency, characteristic of any massive field.

An exact solution to the time-varying problem (\ref{eq:forcedwave}) is thus given by
\begin{equation}\label{eq:phiest1}
	\begin{aligned}
		\tilde{\phi}_\text{wav} &= -\int_{-\infty}^{t} ds\;G(q-q_p(s),s)\frac{be^{-i\omega_cs}}{2i k_c\psi^*(s)}\\
		&=-\int_{-\infty}^{t} ds\;G(q-q_p(t),s)\frac{be^{-i\omega_cs}}{2i k_c\psi^*(t)} \\
		&\qquad\qquad+ \int_{-\infty}^{t} ds\;G(q-q_p(t),s)\frac{be^{-i\omega_cs}}{2i k_c}(\psi^*(t)^{-1}-\psi^*(s)^{-1})\\
		&\qquad\qquad +\int_{-\infty}^{t} ds\;\left(G(q-q_p(t),s) - G(q-q_p(s),s)\right)\frac{be^{-i\omega_cs}}{2i k_c\psi^*(s)}\\
		&=:\phi_{\text{wav},t} + \eps_1(q,t) + \eps_2(q,t),
	\end{aligned}
\end{equation}
where the fixed-time solution $\phi_{\text{wav},t}$ is defined by (\ref{eq:phi_fixed}). The problem thus reduces to bounding $\eps_1$ and $\eps_2$.

Recall from the derivation of the trajectory equation that the particle velocity is quickly dampened to the Bohmian form; in particular, we can bound $u/c = O(U/c)$ with a velocity scale $U$ dependent only on its initial conditions $\dot{q}_{p,0}$ and the Bohmian data $|\psi_0(q_{p,0})|$, $|\nabla\psi_0(q_{p,0})|$. With this in mind, we can choose a time
\[\tau = \tau(|\psi_0(q_{p,0})|,|\nabla\psi_0(q_{p,0})|, |\dot{q}_{p,0}|)\] 
such that the particle remains bounded away from the zero locus of $\psi$ for $t\in[0,\tau]$; say that $\alpha<|\psi(q_p)|<\alpha^{-1}$ for this interval, for a fixed value $\alpha$. Because of the zero Dirichlet conditions, this ensures that $q_p$ remain bounded away from $\partial\Omega$ for $t\leq \tau$.

{Reducing $\tau$ and increasing $U$ as necessary}, we similarly {estimate $|\dot{\psi}(q_p)|= O((mU^2/\hbar)\alpha^{-1})$ over} this interval; this is simply the Schr\"odinger energy scale $E\sim mU^2$, encoded in our non-relativistic hypothesis, put together with our estimate $|\psi(q_p)|<\alpha^{-1}$.

We translate and nondimensionalise the integrand of (\ref{eq:phiest1}) as $s' = \omega_c (s-t)$, as $t_c = 1/\omega_c$ is the dominant timescale in {the Green's function}. Then we expand $\psi$ about $t$ as
\[\psi(s) = \psi(t) + s'd_{s'}\psi + O((s')^2(U^2/c^2)^2\alpha^{-1}),\]
or equivalently,
\[\psi(s)^{-1} = \psi(t)^{-1} - s'\psi^{-2}d_{s'}\psi + O((s')^2(U^2/c^2)^2\alpha^{-1}).\]
With this in mind, we can refine our estimate of $\eps_1$ as follows; writing $r = \|q-q_c\|$, we find
\begin{align*}
	\eps_1 &= \int_{-\infty}^{0} ds' \; G(q-q_p(t),s)\frac{be^{-is'}}{2i k_c\omega_c}(\psi^*(t)^{-1}-\psi^*(s)^{-1})\\
	&= -\int_{-\infty}^{0} ds' \; G(q-q_p(t),s)\frac{be^{-is'}}{2i k_c\omega_c}s'\psi^{-2}d_{s'}\psi + O(\alpha^{-1}U^4/c^4)\\
	&= \frac{bk_c\omega_c}{2\pi}\int_{-\infty}^{0} ds' \; \theta(ct-\|q-q_p(t)\|) \frac{J_1(\omega_c\sqrt{t^2-r^2/c^2})}{\omega_c\sqrt{t^2-r^2/c^2}}\frac{e^{-is'}}{2i k_c\omega_c}s'\psi^{-2}d_{s'}\psi \\
	&\qquad\qquad-\frac{k_c\omega_c r}{4\pi r}\cdot\frac{be^{-ik_cr}d_{s'}\psi}{2ik_c\omega_c\psi^2}+ O(\alpha^{-1}U^4/c^4)\\
	&=b\int_{-\infty}^{0} ds' \; \theta(ct-\|q-q_p(t)\|) \frac{J_1(\omega_c\sqrt{t^2-r^2/c^2})}{\omega_c\sqrt{t^2-r^2/c^2}}\frac{e^{-is'}}{4\pi i }s'\psi^{-2}d_{s'}\psi + O(\alpha^{-1}U^2/c^2)
\end{align*}
Finally, introducing $r' := k_cr$ allows us to calculate
\begin{align*}
	\int_{-\infty}^{0} ds' \; \theta(ct-\|q\|) &\frac{J_1(\omega_c\sqrt{t^2-r^2/c^2})}{\omega_c\sqrt{t^2-r^2/c^2}}\frac{e^{-is'}}{4\pi i }s' \\
	&= \int_{-\infty}^{0} ds' \; \theta(s'-r') \frac{J_1(\sqrt{(s')^2-(r')^2})}{\sqrt{(s')^2-(r')^2}}\frac{e^{-is'}}{4\pi i }s' = O(1),
\end{align*}
as the integral itself converges---to see this, note that the Bessel function decreases as $J_1(x)=O(x^{-1/2})$. But then our estimate for $\psi^{-1}$ gives
\[\eps_1 = O(\alpha^{-1}U^2/c^2),\]
which tells us that feedback from $\phi_\text{wav}$ on $\psi$ vanishes in the non-relativistic limit. A similar argument demonstrates that $\eps_2 = O(\alpha^{-1}U/c)$.

In full, these estimates on $\eps_1$ and $\eps_2$ demonstrate that the error in (\ref{eq:phiforcing}) is at most $O(U/c)$ in the non-relativistic limit, and thus that our ansatz (\ref{eq:ansatz}) for $\phi_\text{wav}$ exactly cancels out the nonlinear wave forcing in this limit. The remaining component $\psi e^{-i\omega_c t}$ thus solves a homogeneous Klein--Gordon equation, and a standard reduction of the resulting unforced Klein--Gordon equation shows that $\psi$ converges to a solution of the free Schr\"odinger equation. 

Putting this field result together with our derivation of the trajectory equation above, we recover the claimed short-time convergence; the long-time convergence follows as discussed at the beginning of this section.

\paragraph*{Note on nonzero potentials.}
As it stands, our result applies to free systems (i.e., $V(q)\equiv 0$) and to systems with rigid walls (i.e., $V(q\in\Omega^c) = +\infty$). For instance, this encompasses several systems previously investigated in the case of walking droplets: the quantum corral (or quantum billiards in general)~\cite{PhysRevE.88.011001}, slit diffraction apparatuses~\cite{PhysRevLett.97.154101,Pucci2018,ellegaard_interaction_2020}, and systems involving sharp walls and beam splitters~\cite{PhysRevA.106.L010203,PhysRevA.108.L060201}.

To adapt our result to the case of nonzero potentials, {we modify} the Klein--Gordon equation as
\[\left(\partial^\mu\partial_\mu + (k_c + V(q)/\hbar c)^2\right)\phi = -\delta^3(q-q_p)\frac{b\gamma^{-1}}{2i k_c\ovl{\phi}^*}.\]
Since the contributions of $V(x)$ appear only at an order $U^2/c^2$ in this equation, the form of $\phi_\text{wav}$ {does not change} from the free case considered here. Moreover, since the Klein--Gordon Green's function changes only at the same relative order, our earlier estimate on $\eps_1$ and $\eps_2$ holds as stated, and our result {goes} through unaffected.

\paragraph*{Note on the choice of $\sigma$.} 
Different expressions for $\sigma$ yield different transient behaviours according to their analogues for (\ref{eq:transient}). As an example, note that the piecewise sawtooth wave
\[\sigma(\theta) = \begin{cases}
\theta - 2\pi & 0 < \theta\leq \pi\\
\theta + 2\pi & \pi < \theta \leq 2\pi
\end{cases}\]
periodically changes the sign of $\sigma/b$ in (\ref{eq:transient}), forcing $u_\text{fluc}$ to jump periodically. In this way, a different choice of $\sigma$ could excite Compton-scale oscillations, better matching the behaviour of the limit studied by Darrow and Bush~\cite{darrow_AM_model}. {We conjecture that $\psi$ still converges to the Schr\"odinger wavefunction with such a coupling, and that the particle oscillates rapidly about a mean Bohmian trajectory.}


\section{Analogue Position Measurements}\label{sec:measure_absolute}
We have demonstrated how, with an appropriate choice of particle-field coupling, the relativistic model of Darrow and Bush~\cite{darrow_AM_model} reduces to traditional Bohmian mechanics in the non-relativistic limit. In this limit, the particle continues to carry a Compton-scale wavepacket, as noted by Darrow and Bush, but this wavepacket no longer radiates energy into the rest of the field.

We can force the system to deviate from Bohmian mechanics by introducing a \emph{thermodynamic} element to the system's evolution, however. In this section, we show how coupling the wave field to a set of heat sinks gives rise to a compelling analogue of wavefunction collapse. 
Along with greater insight into the energetics of the present pilot-wave system, this argument suggests how an experiment with, e.g., the walking droplet system of Couder and Fort~\cite{PhysRevLett.97.154101} might give a macroscopic analogue of quantum measurements.



\begin{figure*}
\centering
\begin{tikzpicture}
	\node[anchor=south west,inner sep=0] (image) at (0,1) {\includegraphics[scale=.18]{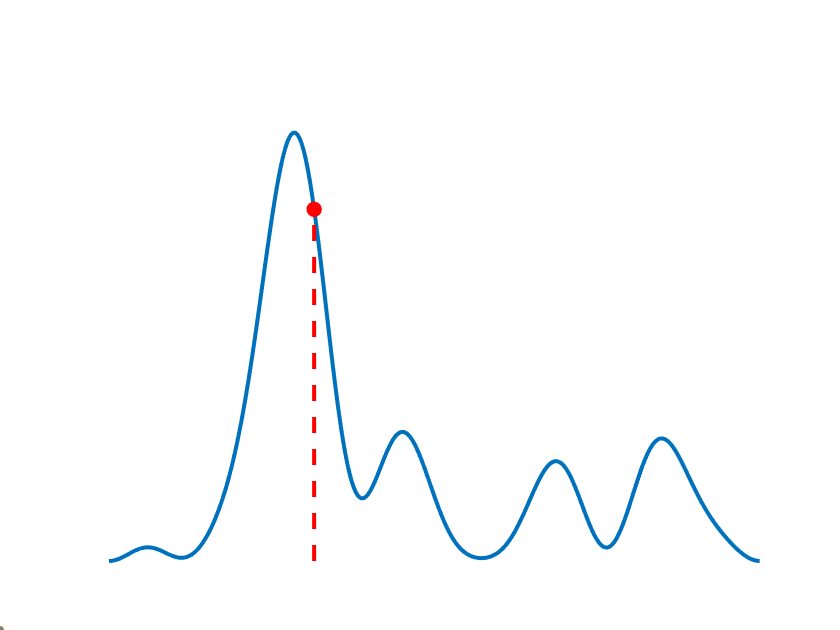}};
	\node[anchor=south west,inner sep=0] (image1) at (4.5,1) {\includegraphics[scale=.18]{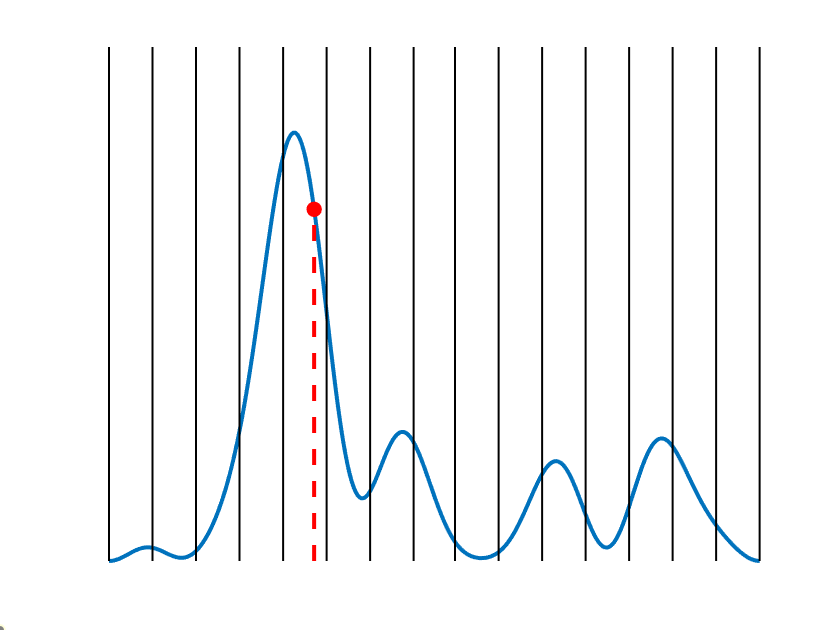}};
	\node[anchor=south west,inner sep=0] (image2) at (9,1) {\includegraphics[scale=.18]{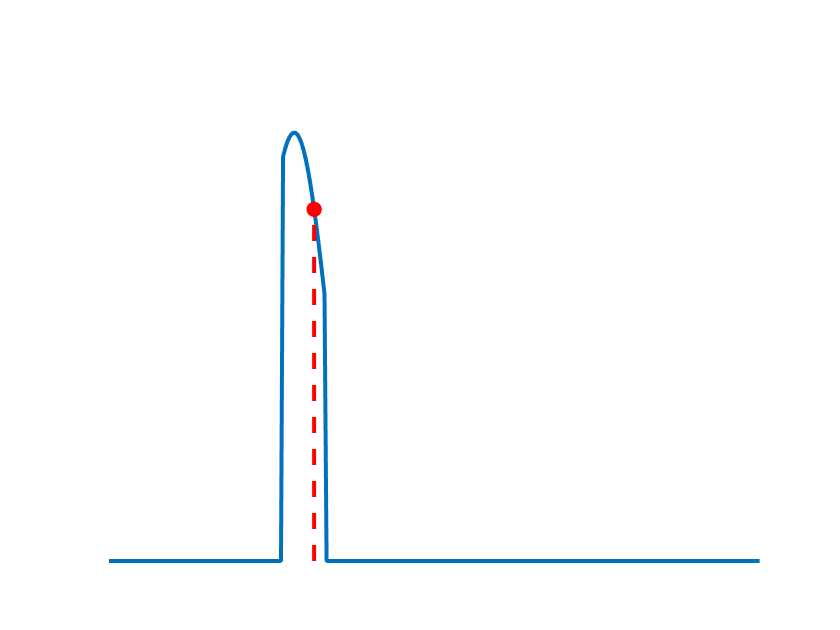}};
	\node[text width=2cm] at (4.9,2) {\Huge$\rightsquigarrow$};
	\node[text width=2cm] at (9.4,2) {\Huge$\rightsquigarrow$};
\end{tikzpicture}
\caption{A depiction of the analogue wave collapse we find in {our} de Broglie-like pilot wave system. In short, we split the domain into measured regions $U_\alpha\subset\mathbb{R}^3$, and we couple each region to a distinct heat sink; this process analogises the energy deposition of a particle into a measurement device. We show that, during such a process, the {wave field} $\psi$ collapses exactly to its projection in the cell containing the particle. We note that there is no conflict with the convergence to Bohmian mechanics we argued in Section \ref{sec:dynamics}; the presence of heat sinks violates the assumptions of those results, at least during the measurement process.}\label{fig:onlyfig}
\end{figure*}

We model a position measurement as follows:
\begin{enumerate}
\item We fix a partition of the domain $\Omega\subset\mathbb{R}^3$ into open cells $U_\alpha\subset\Omega$. During the course of measurement, we enforce that the particle cannot cross from one grid cell to the next, and that field dynamics are decoupled between neighbouring cells.
\item We let the pilot-wave system undergo an energy-minimising flow in each cell, corresponding to a heat sink attached to each cell.
\item Post-measurement, we lift the partition and allow the {system} to continue evolving according to (\ref{eq:oureqsnow}).
\end{enumerate}
The above process is depicted in Fig.~\ref{fig:onlyfig}.

\bigskip
\begin{remark*}
Our analogue of position measurement differs drastically from the usual quantum formalisms, such as that of von Neumann~\cite{Mello_2014}. The primary reason for this is, our results apply only for a single particle\footnote{In fact, a several-particle framework would require nonlocality in order to converge to Bohmian mechanics, as per Bell's theorem~\cite{PhysicsPhysiqueFizika.1.195}. This would sacrifice a key similarity with classical pilot-wave systems, so it is not of present interest.}, so we cannot add new degrees of freedom for the measurement apparatus. As such, we attempt only to model the energy deposition by a single particle as it is measured.
\end{remark*}

\bigskip
To make the measurement process precise, we first derive an expression for the cell-averaged energy.

\paragraph*{Deriving a local mean energy.}

We write
\[\psi_\text{wav} := -\frac{1}{8\pi i k_c\psi^*(q_p)\|q-q_p\|},\]
so that the total field takes the form
\[\phi = (\psi + \psi_\text{wav})e^{-i\omega_ct} =: \psi_\text{tot}e^{-i\omega_ct}.\]
Recall that the function $\psi$ solves the homogeneous Schr\"odinger equation in the non-relativistic limit. The Klein--Gordon energy density takes the form
\[\rho_E(\phi) = mk_c|\partial_t\phi|^2 + mc^2k_c\|\nabla\phi\|^2 + mc^2k_c^3|\phi|^2,\]
which we write in terms of $\psi$ as
\begin{align*}
\rho_E&= mk_c|\partial_t\psi_\text{tot} - i\omega_c\psi_\text{tot}|^2+ mc^2k_c^3\left(|\psi_\text{tot}|^2-|\psi_\text{wav}|^2\right)\\
&\qquad +mc^2k_c\left(\|\nabla\psi_\text{tot}\|^2 - \|\nabla\psi_\text{wav}\|^2\right) + \rho_s/|\psi(q_p)|^2.
\end{align*}
Here, we {have} grouped all terms proportional to $1/|\psi(q_p)|^2$ into the renormalised ``singular energy density'' $\rho_s$; the magnitude of this energy will not affect particle dynamics post-measurement. 

When we integrate over the full cell, cross terms of the form $\psi\nabla^2\psi^*_\text{wav}$ (and thus $\psi^*_\text{wav}\nabla^2\psi$, integrating by parts) can be removed, as the expression
\begin{align*}
\int dV\;\psi\nabla^2\psi^*_\text{wav}
&\propto \int dV\;\frac{\psi}{\psi(q_p)}\delta^3(q-q_p)= 1
\end{align*}
is independent of $\psi$; it corresponds only to an overall shift of energy. Discarding these terms, as well as terms of order $O(U^4/c^4)$, we find a total energy
\begin{align*}
E &= 2mc^2k_c\int dV\;\left(k_c^2|\psi|^2 + \|\nabla\psi\|^2\right)+ \rho_s V/|\psi(q_p)|^2
\\
&\qquad+ \int dV\;\frac{m\omega_c}{4\pi\|q-q_p\|}\left(\op{Re}\frac{\psi\partial_t\psi(q_p)}{\psi(q_p)^2} - \op{Im}\frac{\omega_c\psi}{\psi(q_p)}\right)
\end{align*}
in the cell containing the particle, defining the cell volume $V = \int dV\;1$. The final term {on the right-hand side} approximately vanishes in a short-time average; we can see this rigorously if we assume that the constant contribution of $\psi$ dominates higher-energy terms, {or that $\psi$ is nearly monochromatic.} This leaves us to minimise
\begin{equation}\label{eq:meanenergy}
\begin{aligned}
	\tilde{E} := 2mc^2k_c\int dV\;\left(k_c^2|\psi|^2 + \|\nabla\psi\|^2\right)+ \rho_s V/|\psi(q_p)|^2.
\end{aligned}
\end{equation}

To continue, recall that the Bohmian particle ({which ours converges to}) obeys Born statistics, as
\[P(q_p=q) = \frac{|\psi(q)|^2}{V^{-1}\int dV\; |\psi|^2}.\]
Weighting the mean energy (\ref{eq:meanenergy}) by this distribution, we find a final estimate for the energy in each cell $U_\alpha\subset\Omega$:
\begin{equation}
	\begin{aligned}\label{eq:newenergy}
		\ovl{E}|_{U_\alpha} := \langle \tilde{E}\rangle_{U_\alpha} &= 2mc^2k_c\int dV\;\left(k_c^2|\psi|^2 + \|\nabla\psi\|^2\right) + \mathbf{1}_{q_p\in U_\alpha}\int dV' \frac{\rho_s V^2}{|\psi(q')|^2}\,\frac{|\psi(q')|^2}{\int dV\; |\psi|^2}\\
		&= 2mc^2k_c\int dV\;\left(k_c^2|\psi|^2 + \|\nabla\psi\|^2\right) + \mathbf{1}_{q_p\in U_\alpha}\,\frac{\rho_sV^2}{\int dV\; |\psi|^2}.
	\end{aligned}
\end{equation}
Here, the notation $\mathbf{1}_{q_p\in U_\alpha}$ indicates that the last term is present only if the particle is contained in the cell $U_\alpha$.

\paragraph{Analogue measurements.} Now that we have the energy estimate (\ref{eq:newenergy}) for each cell, we can define our analogue measurement appropriately:

\bigskip
\begin{definition}\label{def:measurement}
Fix a configuration space $\Omega\subseteq\mathbb{R}^3$, and suppose that $(q_p,\psi,\phi_\text{wav})$ evolve according to the non-relativistic dynamics of Section \ref{sec:dynamics}.

A \emph{position measurement} in this system is (a) a collection $U_\alpha\subset\Omega$ of disjoint open sets such that $\Omega\subset \bigcup_\alpha \ovl{U}_\alpha$ and (b) a mean-energy-minimising flow in each cell
\[\psi|_{U_\alpha}\mapsto \psi_t|_{U_\alpha}, \qquad d_t\psi_t|_{U_\alpha} \propto -\nabla_\psi \ovl{E}(\psi_t)|_{U_\alpha},\]
with $\ovl{E}$ as defined in (\ref{eq:newenergy}).
\end{definition}

\bigskip
Importantly, this is a \emph{dynamical} process, and we will see that the system behaviour depends on whether such a process runs for a short or long time. 

The first variation of this quantity, evaluated against a function $h\in L^2(U_\alpha)$, is given by
\begin{align*}
\delta_\psi\ovl{E}(h)|_{U_\alpha} &= 2mc^2k_c\int dV\; \left(k_c^2\psi^*-\nabla^2\psi^*\right)h \\
&\qquad- \frac{\rho_s V^2\mathbf{1}_{q_p\in U_\alpha}}{\left[\int dV\; |\psi|^2\right]^2}\int dV\;\psi^*h,
\end{align*}
which identifies the energy gradient as
\begin{equation}\label{eq:gradient}
\begin{aligned}
	\nabla_\psi\ovl{E}|_{U_\alpha} &= \left(2mc^2k_c^3 - \frac{\rho_s V^2\mathbf{1}_{q_p\in U_\alpha}}{\left[\int dV\; |\psi|^2\right]^2}\right)\psi- 2mc^2k_c\nabla^2\psi.
\end{aligned}
\end{equation}
Now, consider an energy-minimising flow of the form $d_t\psi=-\tilde{\kappa}\nabla_\psi\ovl{E}$, as in Definition \ref{def:measurement}, with $\tilde{\kappa}$ the constant of proportionality. The final term of (\ref{eq:gradient}) is familiar {from the imaginary-time Schr\"odinger equation.} Identifying coefficients with the Schr\"odinger equation suggests writing
\[\tilde{\kappa}=\frac{\kappa\hbar}{4m^2c^2k_c},\]
where $\kappa>0$ is a dimensionless parameter. Then we find
\begin{equation}\label{eq:decay}
\begin{aligned}
	d_t\psi_t = -\kappa\left(\frac{\omega_c}{2} -\frac{(\omega_s/2)\mathbf{1}_{q_p\in U_\alpha}}{V^{-2}\left[\int dV\; |\psi|^2\right]^2}\right)\psi+ \kappa\frac{\hbar}{2m}\nabla^2\psi,
\end{aligned}
\end{equation}
where $\omega_s = \rho_s\hbar/4m^2c^2k_c$ is a system-dependent frequency. If we renormalise $\rho_s$ by assuming a finite radius $O(1/k_c)$ for the particle, then $\omega_s = O(\omega_c)$.

The two terms on the right-hand side of (\ref{eq:decay}) accomplish different tasks, and at different timescales. Over a long timescale $O(\kappa^{-1}\tau_{dB})$, where $\tau_{dB} = \lambda_{dB}/c$, the final term relaxes the system into the Schr\"odinger ground state. This is the expected behaviour of a thermal sink, so it is not surprising that we recover this relaxation.

More importantly, the \emph{first} term in (\ref{eq:decay}) adjusts the normalisation of the wavefunction. Specifically, over the fast timescale $O(\kappa^{-1}\tau_c) = O(\kappa^{-1}\tau_{dB}(U/c)^2)$, the first term scales the wavefunction \emph{without modification} until
\[V^{-1}\int dV\; |\psi|^2 = \sqrt{\frac{2mc^2k_c^3}{\rho_s}} = O(1)\]
in the cell containing the particle. However, any cells $U_\alpha$ for which $q_p\notin U_\alpha$ admit the simpler relaxation
\[d_t\psi_t|_{U_{\alpha}\not\ni q_p} = -\kappa\frac{\omega_c}{2}\psi + \kappa\frac{\hbar}{2m}\nabla^2\psi,\]
which reduces the norm of the wavefunction to zero over the same fast timescale $\kappa^{-1}\tau_c$. This corresponds to the phenomenon of wavefunction collapse in quantum theory: the wavefunction is projected to its component in the cell containing the particle. Since this projection does not affect the particle dynamics in its own cell, the particle is able to evolve post-measurement according to the Bohmian mechanics derived in the preceding section.

\section{Discussion}
{There are several strong similarities between Bohmian mechanics and de Broglie's 1927 proposal~\cite{422ddf4d-944e-34dd-bb65-095bdb4bb08e};} in both, a pilot-wave guides a point-like particle, and Born's probability rule emerges dynamically. {Even still, de Broglie eventually decided that Bohmian mechanics deviated too far from his original picture;} he believed that {the underlying theory should be relativistic, that} the pilot-wave should be a ``physical one... which cannot be arbitrarily normed, and which is distinct from the [Schr\"odinger wavefunction],'' and that the particle must ``form on [the pilot-wave] a small region of high energy concentration, which may be likened... to a moving singularity''~\cite{deBroglie1987}. 

While de Broglie's double-solution program was never able to fully explain quantum mechanics, it provided a striking physical picture that presaged some of the classical pilot-wave systems studied today. These systems---such as the ``walking droplet'' system of Couder and Fort~\cite{PhysRevLett.97.154101} and its various abstractions~\cite{Durey_GPWF,hqfti_i,hqfti_ii,darrow_AM_model}---give rise to a variety of novel, classical behaviours, and they offer dynamical analogues of several key results from quantum theory~\cite{hqa}. However, it remains unclear whether such systems could replicate even single-particle quantum predictions completely.

{Here, we have answered the latter question in the affirmative.} Specifically, we have re-examined the de Broglie-like pilot-wave framework of Darrow and Bush~\cite{darrow_AM_model}, {{in the case} in which a point-particle is coupled to the \emph{phase} of a Klein--Gordon field}. {We have here demonstrated that its non-relativistic limit reduces exactly to Bohmian mechanics, in a highly general setting.} In so doing, we see that de Broglie's double-solution theory---modernised and formalised in the system studied by Darrow and Bush---can be connected {to} Bohmian mechanics using a single Lagrangian framework. Moreover, by pairing this system with an abstracted version of a measuring apparatus, we have shown that it can give rise to a compelling analogue of wavefunction collapse.


The present work further demonstrates the richness of the classical field theory proposed by Darrow and Bush~\cite{darrow_AM_model}, its potential for connecting different classes of pilot-wave systems, {and its potential for capturing new classical quantum analogues.}

\backmatter





\bmhead{Acknowledgments}

I would like to express my sincere gratitude to Professor John Bush (MIT) for his invaluable guidance, support, and encouragement. His constructive dissent has consistently played a crucial role in shaping the direction of my research.

\section*{Declarations}

\subsection*{Funding}

The authors did not receive support from any organization for the submitted work.

\subsection*{Competing Interests}

The authors have no financial or proprietary interests in any material discussed in this article.

\bibliography{Bibliography,HQA2023}


\begin{thebibliography}{33}
\ifx \bisbn   \undefined \def \bisbn  #1{ISBN #1}\fi
\ifx \binits  \undefined \def \binits#1{#1}\fi
\ifx \bauthor  \undefined \def \bauthor#1{#1}\fi
\ifx \batitle  \undefined \def \batitle#1{#1}\fi
\ifx \bjtitle  \undefined \def \bjtitle#1{#1}\fi
\ifx \bvolume  \undefined \def \bvolume#1{\textbf{#1}}\fi
\ifx \byear  \undefined \def \byear#1{#1}\fi
\ifx \bissue  \undefined \def \bissue#1{#1}\fi
\ifx \bfpage  \undefined \def \bfpage#1{#1}\fi
\ifx \blpage  \undefined \def \blpage #1{#1}\fi
\ifx \burl  \undefined \def \burl#1{\textsf{#1}}\fi
\ifx \doiurl  \undefined \def \doiurl#1{\url{https://doi.org/#1}}\fi
\ifx \betal  \undefined \def \betal{\textit{et al.}}\fi
\ifx \binstitute  \undefined \def \binstitute#1{#1}\fi
\ifx \binstitutionaled  \undefined \def \binstitutionaled#1{#1}\fi
\ifx \bctitle  \undefined \def \bctitle#1{#1}\fi
\ifx \beditor  \undefined \def \beditor#1{#1}\fi
\ifx \bpublisher  \undefined \def \bpublisher#1{#1}\fi
\ifx \bbtitle  \undefined \def \bbtitle#1{#1}\fi
\ifx \bedition  \undefined \def \bedition#1{#1}\fi
\ifx \bseriesno  \undefined \def \bseriesno#1{#1}\fi
\ifx \blocation  \undefined \def \blocation#1{#1}\fi
\ifx \bsertitle  \undefined \def \bsertitle#1{#1}\fi
\ifx \bsnm \undefined \def \bsnm#1{#1}\fi
\ifx \bsuffix \undefined \def \bsuffix#1{#1}\fi
\ifx \bparticle \undefined \def \bparticle#1{#1}\fi
\ifx \barticle \undefined \def \barticle#1{#1}\fi
\bibcommenthead
\ifx \bconfdate \undefined \def \bconfdate #1{#1}\fi
\ifx \botherref \undefined \def \botherref #1{#1}\fi
\ifx \url \undefined \def \url#1{\textsf{#1}}\fi
\ifx \bchapter \undefined \def \bchapter#1{#1}\fi
\ifx \bbook \undefined \def \bbook#1{#1}\fi
\ifx \bcomment \undefined \def \bcomment#1{#1}\fi
\ifx \oauthor \undefined \def \oauthor#1{#1}\fi
\ifx \citeauthoryear \undefined \def \citeauthoryear#1{#1}\fi
\ifx \endbibitem  \undefined \def \endbibitem {}\fi
\ifx \bconflocation  \undefined \def \bconflocation#1{#1}\fi
\ifx \arxivurl  \undefined \def \arxivurl#1{\textsf{#1}}\fi
\csname PreBibitemsHook\endcsname

\bibitem[\protect\citeauthoryear{Couder and Fort}{2006}]{PhysRevLett.97.154101}
\begin{barticle}
\bauthor{\bsnm{Couder}, \binits{Y.}},
\bauthor{\bsnm{Fort}, \binits{E.}}:
\batitle{Single-particle diffraction and interference at a macroscopic scale}.
\bjtitle{Phys. Rev. Lett.}
\bvolume{97},
\bfpage{154101}
(\byear{2006})
\doiurl{10.1103/PhysRevLett.97.154101}
\end{barticle}
\endbibitem

\bibitem[\protect\citeauthoryear{Durey and Bush}{2021}]{Durey_GPWF}
\begin{barticle}
\bauthor{\bsnm{Durey}, \binits{M.}},
\bauthor{\bsnm{Bush}, \binits{J.}}:
\batitle{Classical pilot-wave dynamics: The free particle}.
\bjtitle{Chaos}
\bvolume{31},
\bfpage{033136}
(\byear{2021})
\doiurl{10.1063/5.0039975}
\end{barticle}
\endbibitem

\bibitem[\protect\citeauthoryear{Dagan and Bush}{2020}]{hqfti_i}
\begin{barticle}
\bauthor{\bsnm{Dagan}, \binits{Y.}},
\bauthor{\bsnm{Bush}, \binits{J.}}:
\batitle{Hydrodynamic quantum field theory: the free particle}.
\bjtitle{CR Mecanique}
\bvolume{348},
\bfpage{555}--\blpage{571}
(\byear{2020})
\doiurl{10.5802/crmeca.34}
\end{barticle}
\endbibitem

\bibitem[\protect\citeauthoryear{Durey and Bush}{2020}]{hqfti_ii}
\begin{barticle}
\bauthor{\bsnm{Durey}, \binits{M.}},
\bauthor{\bsnm{Bush}, \binits{J.}}:
\batitle{Hydrodynamic quantum field theory: The onset of particle motion and
  the form of the pilot wave}.
\bjtitle{Front. Phys.}
\bvolume{8},
\bfpage{300}
(\byear{2020})
\doiurl{10.3389/fphy.2020.00300}
\end{barticle}
\endbibitem

\bibitem[\protect\citeauthoryear{Darrow and Bush}{2024}]{darrow_AM_model}
\begin{botherref}
\oauthor{\bsnm{Darrow}, \binits{D.}},
\oauthor{\bsnm{Bush}, \binits{J.W.M.}}:
Revisiting de {Broglie's} double-solution pilot-wave theory with a
  {Lorentz}-covariant {Lagrangian} framework.
Symmetry
\textbf{16}(2)
(2024)
\end{botherref}
\endbibitem

\bibitem[\protect\citeauthoryear{de~Broglie}{1956}]{deBroglie1956}
\begin{bbook}
\bauthor{\bsnm{Broglie}, \binits{L.}}:
\bbtitle{Une Tentative D'interpr\'{e}tation Causale et Nonlin\'{e}aire de la
  M\'{e}canique Ondulatoire: la Th\'{e}orie de la Double solution}.
\bpublisher{Gautier-Villars},
\blocation{Paris}
(\byear{1956})
\end{bbook}
\endbibitem

\bibitem[\protect\citeauthoryear{de~Broglie}{1970}]{deBroglie1970}
\begin{barticle}
\bauthor{\bsnm{Broglie}, \binits{L.}}:
\batitle{The reinterpretation of wave mechanics}.
\bjtitle{Found. Phys.}
\bvolume{1}(\bissue{1}),
\bfpage{5}--\blpage{15}
(\byear{1970})
\end{barticle}
\endbibitem

\bibitem[\protect\citeauthoryear{{de Broglie}}{1925}]{1925AnPh...10...22D}
\begin{barticle}
\bauthor{\bsnm{{de Broglie}}, \binits{L.}}:
\batitle{Recherches sur la th{\'e}orie des quanta}.
\bjtitle{Ann. Phys.}
\bvolume{10}(\bissue{3}),
\bfpage{22}--\blpage{128}
(\byear{1925})
\doiurl{10.1051/anphys/192510030022}
\end{barticle}
\endbibitem

\bibitem[\protect\citeauthoryear{Bacciagaluppi and
  Valentini}{2009}]{bacciagaluppi2009quantum}
\begin{bbook}
\bauthor{\bsnm{Bacciagaluppi}, \binits{G.}},
\bauthor{\bsnm{Valentini}, \binits{A.}}:
\bbtitle{Quantum Theory at the Crossroads: Reconsidering the 1927 Solvay
  Conference}.
\bpublisher{Cambridge University Press}, \blocation{???}
(\byear{2009})
\end{bbook}
\endbibitem

\bibitem[\protect\citeauthoryear{Bohm}{1952a}]{Bohm1952a}
\begin{barticle}
\bauthor{\bsnm{Bohm}, \binits{D.}}:
\batitle{A suggested interpretation of the quantum theory in terms of hidden
  variables, {I}}.
\bjtitle{Phys. Rev.}
\bvolume{85},
\bfpage{66}--\blpage{179}
(\byear{1952})
\end{barticle}
\endbibitem

\bibitem[\protect\citeauthoryear{Bohm}{1952b}]{Bohm1952b}
\begin{barticle}
\bauthor{\bsnm{Bohm}, \binits{D.}}:
\batitle{A suggested interpretation of the quantum theory in terms of hidden
  variables, {II}}.
\bjtitle{Phys. Rev.}
\bvolume{85},
\bfpage{180}--\blpage{193}
(\byear{1952})
\end{barticle}
\endbibitem

\bibitem[\protect\citeauthoryear{von Neumann and
  Beyer}{2018}]{30573279-e8ec-3f1e-b0ba-eaf73275f821}
\begin{bbook}
\bauthor{\bsnm{Neumann}, \binits{J.}},
\bauthor{\bsnm{Beyer}, \binits{R.T.}}:
\bbtitle{Mathematical Foundations of Quantum Mechanics: New Edition}.
\bpublisher{Princeton University Press},
\blocation{Princeton, NJ}
(\byear{2018}).
\burl{http://www.jstor.org/stable/j.ctt1wq8zhp}
Accessed 2024-01-11
\end{bbook}
\endbibitem

\bibitem[\protect\citeauthoryear{Holland}{1993}]{Holland1993}
\begin{bbook}
\bauthor{\bsnm{Holland}, \binits{P.R.}}:
\bbtitle{The Quantum Theory of Motion: An Account of the de Broglie-Bohm Causal
  Interpretation of Quantum Mechanics}.
\bpublisher{Cambridge University Press},
\blocation{Cambridge, U.K.}
(\byear{1993})
\end{bbook}
\endbibitem

\bibitem[\protect\citeauthoryear{de~Broglie}{1952}]{422ddf4d-944e-34dd-bb65-095bdb4bb08e}
\begin{barticle}
\bauthor{\bsnm{Broglie}, \binits{L.}}:
\batitle{La physique quantique restera-t-elle indéterministe?}
\bjtitle{Rev. Hist. Sci.}
\bvolume{5}(\bissue{4}),
\bfpage{289}--\blpage{311}
(\byear{1952}).
Accessed 2024-03-08
\end{barticle}
\endbibitem

\bibitem[\protect\citeauthoryear{Bell}{1966}]{RevModPhys.38.447}
\begin{barticle}
\bauthor{\bsnm{Bell}, \binits{J.S.}}:
\batitle{On the problem of hidden variables in quantum mechanics}.
\bjtitle{Rev. Mod. Phys.}
\bvolume{38},
\bfpage{447}--\blpage{452}
(\byear{1966})
\doiurl{10.1103/RevModPhys.38.447}
\end{barticle}
\endbibitem

\bibitem[\protect\citeauthoryear{Bell}{1964}]{PhysicsPhysiqueFizika.1.195}
\begin{barticle}
\bauthor{\bsnm{Bell}, \binits{J.S.}}:
\batitle{On the {Einstein--Podolsky--Rosen} paradox}.
\bjtitle{Phys. Phys. Fiz.}
\bvolume{1},
\bfpage{195}--\blpage{200}
(\byear{1964})
\doiurl{10.1103/PhysicsPhysiqueFizika.1.195}
\end{barticle}
\endbibitem

\bibitem[\protect\citeauthoryear{Aspect et~al.}{1982}]{PhysRevLett.49.1804}
\begin{barticle}
\bauthor{\bsnm{Aspect}, \binits{A.}},
\bauthor{\bsnm{Dalibard}, \binits{J.}},
\bauthor{\bsnm{Roger}, \binits{G.}}:
\batitle{Experimental test of {Bell's} inequalities using time-varying
  analyzers}.
\bjtitle{Phys. Rev. Lett.}
\bvolume{49},
\bfpage{1804}--\blpage{1807}
(\byear{1982})
\doiurl{10.1103/PhysRevLett.49.1804}
\end{barticle}
\endbibitem

\bibitem[\protect\citeauthoryear{Sutherland}{2019a}]{sutherland2019incorporating}
\begin{botherref}
\oauthor{\bsnm{Sutherland}, \binits{R.}}:
Incorporating action and reaction into a particle interpretation for quantum
  mechanics -- {Dirac} case
(2019)
\end{botherref}
\endbibitem

\bibitem[\protect\citeauthoryear{Sutherland}{2019b}]{sutherland2019incorporating2}
\begin{botherref}
\oauthor{\bsnm{Sutherland}, \binits{R.}}:
Incorporating action and reaction into a particle interpretation for quantum
  mechanics -- {Schr\"odinger} case
(2019)
\end{botherref}
\endbibitem

\bibitem[\protect\citeauthoryear{Bush and Oza}{2021}]{hqa}
\begin{botherref}
\oauthor{\bsnm{Bush}, \binits{J.W.M.}},
\oauthor{\bsnm{Oza}, \binits{A.U.}}:
Hydrodynamic quantum analogs.
Rep. Prog. Phys.
\textbf{84}(1)
(2021)
\end{botherref}
\endbibitem

\bibitem[\protect\citeauthoryear{Pucci et~al.}{2018}]{Pucci2018}
\begin{barticle}
\bauthor{\bsnm{Pucci}, \binits{G.}},
\bauthor{\bsnm{Harris}, \binits{D.M.}},
\bauthor{\bsnm{Faria}, \binits{L.M.}},
\bauthor{\bsnm{Bush}, \binits{J.W.M.}}:
\batitle{Walking droplets interacting with single and double slits}.
\bjtitle{J. Fluid Mech.}
\bvolume{835},
\bfpage{1136}--\blpage{1156}
(\byear{2018})
\end{barticle}
\endbibitem

\bibitem[\protect\citeauthoryear{Ellegaard and
  Levinsen}{2020}]{ellegaard_interaction_2020}
\begin{barticle}
\bauthor{\bsnm{Ellegaard}, \binits{C.}},
\bauthor{\bsnm{Levinsen}, \binits{M.T.}}:
\batitle{Interaction of wave-driven particles with slit structures}.
\bjtitle{Physical Review E}
\bvolume{102}(\bissue{2}),
\bfpage{023115}
(\byear{2020})
\doiurl{10.1103/PhysRevE.102.023115} .
Accessed 2021-08-27
\end{barticle}
\endbibitem

\bibitem[\protect\citeauthoryear{Fort et~al.}{2010}]{fort_2010}
\begin{botherref}
\oauthor{\bsnm{Fort}, \binits{E.}},
\oauthor{\bsnm{Eddi}, \binits{A.}},
\oauthor{\bsnm{Boudaoud}, \binits{A.}},
\oauthor{\bsnm{Moukhtar}, \binits{J.}},
\oauthor{\bsnm{Couder}, \binits{Y.}}:
Path-memory induced quantization of classical orbits.
P. Natl. Acad. Sci. USA
\textbf{107}
(2010)
\doiurl{10.1073/pnas.1007386107}
\end{botherref}
\endbibitem

\bibitem[\protect\citeauthoryear{Harris and Bush}{2014}]{harris_bush_2014}
\begin{barticle}
\bauthor{\bsnm{Harris}, \binits{D.M.}},
\bauthor{\bsnm{Bush}, \binits{J.W.M.}}:
\batitle{Droplets walking in a rotating frame: from quantized orbits to
  multimodal statistics}.
\bjtitle{J. Fluid Mech.}
\bvolume{739},
\bfpage{444}--\blpage{464}
(\byear{2014})
\doiurl{10.1017/jfm.2013.627}
\end{barticle}
\endbibitem

\bibitem[\protect\citeauthoryear{Perrard et~al.}{2014}]{perrard_2014}
\begin{barticle}
\bauthor{\bsnm{Perrard}, \binits{S.}},
\bauthor{\bsnm{Labousse}, \binits{M.}},
\bauthor{\bsnm{Miskin}, \binits{M.}},
\bauthor{\bsnm{Fort}, \binits{E.}},
\bauthor{\bsnm{Couder}, \binits{Y.}}:
\batitle{Self-organization into quantized eigenstates of a classical
  wave-driven particle}.
\bjtitle{Nat. Commun.}
\bvolume{5},
\bfpage{3219}
(\byear{2014})
\doiurl{10.1038/ncomms4219}
\end{barticle}
\endbibitem

\bibitem[\protect\citeauthoryear{Oza et~al.}{2018}]{spinstates}
\begin{barticle}
\bauthor{\bsnm{Oza}, \binits{A.}},
\bauthor{\bsnm{Rosales}, \binits{R.}},
\bauthor{\bsnm{Bush}, \binits{J.W.M.}}:
\batitle{Hydrodynamic spin states}.
\bjtitle{Chaos: An Interdisciplinary Journal of Nonlinear Science}
\bvolume{28},
\bfpage{096106}
(\byear{2018})
\doiurl{10.1063/1.5034134}
\end{barticle}
\endbibitem

\bibitem[\protect\citeauthoryear{Harris et~al.}{2013}]{PhysRevE.88.011001}
\begin{barticle}
\bauthor{\bsnm{Harris}, \binits{D.M.}},
\bauthor{\bsnm{Moukhtar}, \binits{J.}},
\bauthor{\bsnm{Fort}, \binits{E.}},
\bauthor{\bsnm{Couder}, \binits{Y.}},
\bauthor{\bsnm{Bush}, \binits{J.W.M.}}:
\batitle{Wavelike statistics from pilot-wave dynamics in a circular corral}.
\bjtitle{Phys. Rev. E}
\bvolume{88},
\bfpage{011001}
(\byear{2013})
\doiurl{10.1103/PhysRevE.88.011001}
\end{barticle}
\endbibitem

\bibitem[\protect\citeauthoryear{Frumkin et~al.}{2022}]{PhysRevA.106.L010203}
\begin{barticle}
\bauthor{\bsnm{Frumkin}, \binits{V.}},
\bauthor{\bsnm{Darrow}, \binits{D.}},
\bauthor{\bsnm{Bush}, \binits{J.W.M.}},
\bauthor{\bsnm{Struyve}, \binits{W.}}:
\batitle{Real surreal trajectories in pilot-wave hydrodynamics}.
\bjtitle{Phys. Rev. A}
\bvolume{106},
\bfpage{010203}
(\byear{2022})
\doiurl{10.1103/PhysRevA.106.L010203}
\end{barticle}
\endbibitem

\bibitem[\protect\citeauthoryear{Schoene}{1979}]{SCHOENE197936}
\begin{barticle}
\bauthor{\bsnm{Schoene}, \binits{A.Y.}}:
\batitle{On the nonrelativistic limits of the {Klein--Gordon} and {Dirac}
  equations}.
\bjtitle{Journal of Mathematical Analysis and Applications}
\bvolume{71}(\bissue{1}),
\bfpage{36}--\blpage{47}
(\byear{1979})
\doiurl{10.1016/0022-247X(79)90216-6}
\end{barticle}
\endbibitem

\bibitem[\protect\citeauthoryear{}{Accessed 2023}]{wolfram_greens}
\begin{botherref}
Green's function for the {Klein--Gordon} equation.
{Wolfram Research}
(Accessed 2023).
\url{functions.wolfram.com/Bessel-TypeFunctions/BesselJ/31/02/}
\end{botherref}
\endbibitem

\bibitem[\protect\citeauthoryear{Frumkin and Bush}{2023}]{PhysRevA.108.L060201}
\begin{barticle}
\bauthor{\bsnm{Frumkin}, \binits{V.}},
\bauthor{\bsnm{Bush}, \binits{J.W.M.}}:
\batitle{Misinference of interaction-free measurement from a classical system}.
\bjtitle{Phys. Rev. A}
\bvolume{108},
\bfpage{060201}
(\byear{2023})
\doiurl{10.1103/PhysRevA.108.L060201}
\end{barticle}
\endbibitem

\bibitem[\protect\citeauthoryear{Mello}{2014}]{Mello_2014}
\begin{bchapter}
\bauthor{\bsnm{Mello}, \binits{P.A.}}:
\bctitle{The von {Neumann} model of measurement in quantum mechanics}.
In: \bbtitle{AIP Conference Proceedings}.
\bpublisher{AIP Publishing LLC},
\blocation{Melville, NY}
(\byear{2014}).
\doiurl{10.1063/1.4861702} .
\burl{http://dx.doi.org/10.1063/1.4861702}
\end{bchapter}
\endbibitem

\bibitem[\protect\citeauthoryear{{de Broglie}}{1987}]{deBroglie1987}
\begin{barticle}
\bauthor{\bsnm{{de Broglie}}, \binits{L.}}:
\batitle{Interpretation of quantum mechanics by the double solution theory}.
\bjtitle{Ann. Fond. Louis de Broglie}
\bvolume{12},
\bfpage{1}--\blpage{23}
(\byear{1987})
\end{barticle}
\endbibitem

\end{thebibliography}

\end{document}